\documentclass[preprint,pra,showpacs,nofootinbib]{revtex4}

\usepackage{amsfonts,amsmath,amssymb,amsthm}
\usepackage{latexsym}
\usepackage{bbm,bm}
\usepackage{graphicx}

\newcommand{\uv}{{\bm u}}
\newcommand{\rv}{{\bm r}}
\newcommand{\ra}{\rangle}
\newcommand{\la}{\langle}
\newcommand{\vp}{\varphi}
\newcommand{\tr}{\mathrm{Tr}}

\begin{document}

\title{Maximally entangled three-qubit states via
geometric measure of entanglement}
\author{Sayatnova Tamaryan}
\affiliation{Theory Department, Yerevan Physics Institute,
Yerevan, 375036, Armenia}
\author{Tzu-Chieh Wei}
\affiliation{Institute for Quantum Computing and Department of Physics and
Astronomy, University of Waterloo, Waterloo, Ontario, Canada} \affiliation{
Department of Physics and Astronomy, University of British Columbia,
Vancouver, British Columbia, Canada}
\author{DaeKil Park}
\affiliation{Department of Physics, Kyungnam University, Masan,
631-701, Korea}

\begin{abstract}
Bipartite maximally entangled states have the property that the largest
Schmidt coefficient reaches its lower bound. However, for multipartite states
the standard Schmidt decomposition generally does not exist. We use a
generalized Schmidt decomposition and the geometric measure of entanglement to
characterize three-qubit {\it pure\/} states and derive a single-parameter
family of maximally entangled three-qubit states.  The paradigmatic
Greenberger-Horne-Zeilinger (GHZ) and W states emerge as extreme members in
this family of maximally entangled states. This family of states possess
different trends of entanglement behavior: in going from GHZ to W states the
geometric measure, the relative entropy of entanglement, and the bipartite
entanglement all increase monotonically whereas the three-tangle and
bi-partition negativity both decrease monotonically.
\end{abstract}

\pacs{03.67.Mn, 03.65.Ud}

\maketitle

\section{Introduction}
Maximally entangled states are in essence the natural units of entanglement
with which one would like to compare all quantum states. A well-motivated
approach to compare different entangled states and quantify their entanglement
is to consider how they can transform to each other under local operations and
classical communications (LOCC) in the asymptotic regime. The main question is
then to quantify the optimal rate of conversion between two given
states~\cite{ben}. For bipartite systems this gives rise to the two basic
operational entanglement measures: the entanglement cost (EC)~\cite{ben,woot}
and the distillable entanglement (ED)~\cite{ben,dist-2}, with Bell
states~\cite{epr,bel} emerging as the standard metric of entanglement. While
the latter measure is the rate at which copies of the maximally entangled
state can be concentrated from those of a given state, the former is the rate
at which copies of the maximally entangled state need to be consumed
for the preparation of the given state~\cite{dist-2,woot,vid-cir,hor}.

In contrast, there is no simple and unique characterization of a maximally
entangled state in multi-partite settings.  It has been a long standing
question whether there exists a finite minimal reversible entanglement
generating set (MREGS)~\cite{ben-schm}, as the states in MREGS would provide
several distinct metrics of entanglement and hence the generalization of ED
and EC would become possible. In addition to the issue of interconversion in
the asymptotic limit, another challenge lies in the fact that multi-partite
states can be entangled in several inequivalent ways~\cite{w,four} and that
the number of these likely grows exponentially with the number of
parties~\cite{four}.

 Perhaps, to  get to certain handle of these problems it is useful, as an
initial stage, to elucidate an important question of which pure states can be
regarded as the maximally entangled states~\cite{pure-05}. A clear definition
of these states and the search of an effective method for deriving them could
offer a step towards understanding multipartite entanglement, including the
structure of the Hilbert space and multipartite entanglement measures. Any
multipartite entanglement measure is perhaps a likely starting point for such
a definition of maximally entangled states, as for each measure there must
exist a set of states which are maximally entangled. However, these states may
not be maximally entangled using a different measure. Therefore, one then has
to choose an entanglement measure that gives rise to a set of maximally
entangled states that include the known ones in the set. Furthermore, the
selected measure should be suitable for any number of parties with any
dimensions, in order for the notion of maximal entanglement be properly
quantified.

In the setting of three qubits, the GHZ state $|{\rm GHZ}\rangle\equiv
(|000\rangle+|111\rangle)/\sqrt{2}$~\cite{ghz} and the W state $|W\rangle
\equiv (|001\rangle +|010\rangle +|100\rangle)/\sqrt{3}$~\cite{w} have been
realized as  two inequivalent entangled states that cannot be transformed to
each other via LOCC nor even stochastically. GHZ state seems to be the most
natural generalization of Bell states and possesses a maximum tri-partite
entanglement characterized by the three-tangle~\cite{coff}. On the other hand,
W state possesses zero three-tangle and yet maximizes the residual bi-partite
entanglement~\cite{w}. In some sense, both states are maximally entangled.
However, the exact value of entanglement depends on the choice of entanglement
measures, and it is likely that different measures may give different ordering
of the entanglement quantity. For example, using the relative entropy of
entanglement (ER)~\cite{er}, one has $E_R({\rm GHZ})=1$ and $E_R({\rm
W})=\log_2(9/4)>1$~\cite{erw}. However, using the negativity $N$ across any
bi-partition, one obtains $N({\rm GHZ})=1 > N({\rm
W})=2\sqrt{2}/3$~\cite{neg}.  Hence, the notion of being maximally entangled
can depend on the choice of entanglement measures. In the setting of two
qubits, in contrast, the Bell states emerge as the ones that possess maximal
entanglement, independent of the choice of measures. Nonetheless, the
dependence does appear when one consider two-qubit mixed states, and the form
of maximally entangled two-qubit mixed states~\cite{2qubit1,2qubit2,2qubit3}
(e.g., parametrized by entropy) can actually vary~\cite{2qubit3}.

 Here we characterize maximal entanglement of three-qubit {\it pure\/} states
via the geometric measure of entanglement~\cite{wei}. Under this measure, the
W state turns out to possess the maximal entanglement. To see whether the GHZ
can fit into our picture, we investigate the maximal entanglement with a
single parameter $\gamma$, which we call the ``gauge'' phase (which appears in
the generalized Schmidt decomposition~\cite{hig,acin,gsd} and to be defined
below), in analogue to the two-qubit maximally entangled mixed states
parametrized by the entropy. We derive the whole family of maximally entangled
states (characterized by the gauge phase) and show that both GHZ and W emerge
as maximally entangled states at two different gauge phases, at the opposite
ends of this family.  Interestingly, in going from GHZ to W as the gauge phase
increases, the geometric measure and the relative entropy of entanglement and
the bipartite entanglement all increase monotonically whereas the three-tangle
and bipartite negativity both decrease monotonically (see Fig.~\ref{fig:5}).
The two different trends of these entanglement monotones for this family of
states imply that no states in the family can be interconverted
deterministically to each other via LOCC using a single copy~\cite{vidal}.
Furthermore, our analysis of using three-tangle shows that any states can be
probabilistically converted to one another via stochastic LOCC in the family
except to/from the W state. This is because the whole family of the maximally
entangled three-qubit states, except the W state, belong to the GHZ-class, a
classification introduced by D\"ur, Vidal and Cirac~\cite{vid-cir}.

The paper is organized as follows. In Sec. II we use a generalized Schmidt
decomposition to parameterize three-qubit states. In Sec. III we consider the
case of vanishing gauge phase and obtain the GHZ-state. In Sec. IV we consider
the case of maximal gauge phase and obtain the W-state. In Sec. V we consider
the general case and derive an one-parameter  family of maximally entangled
states. Furthermore, we discuss several entanglement properties and
interconversion of states in the family of the maximally entangled states. In
Sec. VI we make concluding remarks.

\section{Generalized Schmidt Decompositions}

In general one needs fourteen real parameters to describe a three-qubit pure
state. Carteret et al.~\cite{hig} and Ac\'in et al.~\cite{acin} have
independently proposed generalizations of the Schmidt decomposition in
multipartite settings (three qubits, in particular) to reduce the number of
necessary parameters. The generalized Schmidt decomposition (GSD) that we
shall use is a variant~\cite{gsd} that is closely related to the geometric
measure, and hence it is more appropriate to our discussions here. Let us
briefly discuss this decomposition. For any multi-qubit pure state, one can
always search a closest product state to it, and the local states of the
closest product state and their orthonormal states uniquely (up to phases)
determine the local bases one uses to express the multi-qubit state. One can
also relabel the closest product as $|000\rangle$ and the local state
orthonormal to $|0\rangle$ by $|1\rangle$, and arrive at the following
expression~\cite{gsd}:
\begin{equation}\label{int.gsd}
|\psi\ra=g|000\ra+t_1|011\ra+t_2|101\ra+t_3|110\ra+
e^{i\gamma}h|111\ra,
\end{equation}
where the labels within each ket refer to qubits A, B and C (or 1, 2 and 3) in
that order and will be suppressed whenever no confusion occurs. Furthermore,
the parameters in the decomposition satisfy
\begin{equation}\label{para.gsd}
g\ge t_i,h\ge0, \ -\pi/2\le\gamma\le\pi/2, \ \mbox{and} \,
g^2+h^2+t_1^2+t_2^2+t_3^2=1.
\end{equation}
 The reason that we do not have
$|001\rangle$, $|010\rangle$, and $|100\rangle$ is that the component
$|000\rangle$ is the closest product state and if there were any of the three
other components, one could have absorbed them and increased the maximal
overlap $g$. We shall refer to $\gamma$ the ``gauge phase'', as such a factor
will necessarily appear in any such decomposition but it may be associated
with $|111\rangle$ or with $|100\rangle$, as in Ref.~\cite{acin}. In what
follows we shall analyze only positive values of the gauge phase since the
maximal overlap $g$ is an even function on $\gamma$.

 Why do we choose to parametrize the family of maximally entangled states
by $\gamma$ (or equivalently the phase in Ref.~\cite{acin})? Consider the
unitary three-qubit gate: control-control-phase  (CCP) gate  that multiplies
the computational state $|111\rangle$ by a phase factor $e^{i \phi}$ but
leaves unchanged the remaining seven basis states
$|000\rangle,|001\rangle,\dots|110\rangle$. This gate can be used to generate
entanglement (one says that it is an entangling gate). If we start with the
(un-normalized) product state
$(|0\rangle+|1\rangle)(|0\rangle+|1\rangle)(|0\rangle+|1\rangle)$ and apply to
it the CCP gate, one obtain
$|000\rangle+|001\rangle+|010\rangle+|100\rangle+|110\rangle+|101\rangle+|011\rangle+
e^{i \phi} |111\rangle$. It can be shown that this state is entangled and has
a three-tangle $\tau=|\sin \phi|/8$ and, moreover, the reduced two-qubit state
can also be entangled, depending on $\phi$. The states in the generalized
Schmidt form~(\ref{int.gsd}) with fixed $g,t_1,t_2,t_3,h$ but different
$\gamma$'s can be connected by the entangling CCP gate with an appropriate
value of phase $\phi$.
 In contrast,
any local unitary transformation that changes any of the five magnitude
parameters will generally change the values of the others and most likely take
the state out of the generalized Schmidt form. Since any three-qubit pure
state can be made into this form by local unitary transformations, it seems
natural to take the gauge phase $\gamma$ as the parameter to search for the
family of maximally entangled states.

 The decomposition into the form~(\ref{int.gsd}) captures all the three-qubit
pure states, up to local unitary transformations. It is thus a convenient
starting point for searching maximally entangled three-qubit states. When
$|000\ra$ is the closest product state and the parameters satisfy
condition~(\ref{para.gsd}), we shall refer to the decomposition as being in
the canonical form. In this canonical form, the nearest product state
$|000\ra$ is a stationary point for $\psi$ and should satisfy the stationarity
equations~\cite{hig,wei}, which represent a nonlinear eigenvalue problem:
\begin{equation}\label{int.stat}
\la i_1i_2|\psi\ra=\mu_i|i_3\ra,\, \la
i_1i_3|\psi\ra=\mu_i|i_2\ra,\, \la i_2i_3|\psi\ra=\mu_i|i_1\ra,
\end{equation}
where we can always restrict ourselves to $\mu_i\geq0$ by adjusting phases of
local states $|i\ra$.  The above stationary conditions arise from the
requirement that the overlap with product states be extremal under the
constraint that product states be normalized; for derivation, see
Refs.~\cite{wei,sud}. The resulting equations generalize the linear eigenvalue
problem to a nonlinear form, which has many features different from the linear
scenario~\cite{sud}. (When the number of parties is two, the generally
nonlinear eigenvalue equation becomes linear.) The largest Schmidt coefficient
is the maximal nonlinear eigenvalue, i.e. $g=\max\limits_i(\mu_i)$, and thus
the nearest product state is the dominant eigenvector of stationarity
equations. As mentioned earlier, it uniquely defines the factorizable basis of
GSD consisting of $|000\ra$ and its complimentary orthogonal product
states~\cite{gsd}.

Our goal is to find a family of maximally entangled states, parameterized by
the gauge phase $\gamma$. First, we observe that the coefficient $g$ in
the canonical decomposition measures the overlap (or the angle) from
$\psi$ to the closest unentangled state~\cite{wei,wern} and cannot decrease
under LOCC~\cite{vedr,bno}. Hence it should be minimal for the maximally
entangled states for all other possible parameters ($t$ and $h$) given fixed
gauge phase $\gamma$. The crucial point then lies in finding the lower bound
on $g$~\cite{guh}. In the case of generic three-qubit states stationarity
equations have six solutions. Let $\mu_1$ be the largest eigenvalue and hence
$g=\mu_1$. Then clearly we have
\begin{equation}\label{int.low}
g\geq\mu=\max(\mu_2,\mu_3,...,\mu_6).
\end{equation}
This is a strong lower bound on $g$. If one can compute all the eigenvalues
$\mu_i$, then one can find the lowest value of $g$ directly. We shall first do
this for special cases $\gamma=0$ and $\gamma=\pi/2$, and the celebrated GHZ
and W states emerge as the maximally entangled states, respectively.

In general, the derivation of eigenvalues $\mu_i$ gives rise to unsolvable
equations~\cite{shared,sud}. Fortunately, there is a realizable method that
gives the desired lower bound. The essence of the method is the following. If
the three-qubit state is maximally entangled, then the above inequality would
be saturated. Consequently $g$ must coincide with $\mu$ and thus the largest
eigenvalue of stationarity equations should be degenerated. This requirement
imposes a condition (i.e., {\it degeneracy condition}) on state parameters,
which can then be deduced from stationarity equations. An example is the GHZ
state $(|000\ra+|111\ra)/\sqrt{2}$, where $|000\ra$ and $|111\ra$ are two such
``degenerate'' states with $\mu=1/\sqrt{2}$. Following the above procedure we
derive the degeneracy condition for three-qubit states and single out states
satisfying this condition. Next we find among these states the one with the
minimal $g$ (over the remaining free parameters) for a given value of the
gauge phase.

We have seen that the GHZ and W states emerge as maximally entangled states in
different contexts, such as via three-tangle and residual bipartite
entanglement, respectively, and they are invariant under permuting parties. It
is thus natural to assume that maximally entangled states can be made
symmetric. This ansatz will be verified against numerical experiments (see
Fig.~\ref{fig:random}). Thus, we shall work with the assumption that
$t_1=t_2=t_3\equiv t$ and consider from now on states of the form
\begin{equation}\label{int.gsd2}
|\psi\ra=g|000\ra+t|011\ra+t|101\ra+t|110\ra+ e^{i\gamma}h|111\ra.
\end{equation}

Eigenvalues $\mu_i$ of the state Eq.(\ref{int.gsd2}) satisfy a polynomial
equation of degree 12 and are roots of the characteristic
polynomial~\cite{sud}. Local states $|i\ra$ satisfy the analogous polynomial
equation and the general case remains to some extent intractable. However, at
extreme values of the gauge phase $\gamma=0$ and $\gamma=\pi/2$ this
polynomial equation can be factorized to cubic equations. Another major step
towards analytic solutions is the following. Each of these cubic equations can
be further factorized to a linear and quadratic equations. This observation
allows us to find all roots of the characteristic polynomial. Some of them
have no associated eigenvectors and hence are irrelevant. Some others never
maximize the overlap since their value is smaller than $\max(h,t)$ in whole
state parameter space~\cite{minimum}. Remaining solutions are listed in the
Appendix.

\section{GHZ state}
 Consider first
the case $\gamma=0$. All of these states are symmetric and have real
coefficients
\begin{equation}\label{ghz.gsd}
|\psi_0\ra=g|000\ra+t(|011\ra+|101\ra+|110\ra)+h|111\ra.
\end{equation}
Owing to these properties the eigenvector with eigenvalue $\mu$ is
symmetric, i.e. has a form $|qqq\ra$, and its constituents $|q\ra$
have real coefficients. In this reason we will derive here only
symmetric solutions as the asymmetric ones give strictly $\mu<g$.

We parameterize the pure one-qubit state $|q\ra$ by a single angle
$|q\ra=\cos\theta|0\ra+\sin\theta|1\ra$ and insert it into
Eq.~(\ref{int.stat}). The result is a pair of equations for unknowns $\theta$
and $\mu$
\begin{equation}\label{gz.eq}
g\cos^2\theta+t\sin^2\theta=\mu\cos\theta,\quad
h\sin^2\theta+t\sin2\theta=\mu\sin\theta.
\end{equation}
These equations have an obvious solution $\mu_1=g,\cos\theta=1$ reflecting the
fact that the state is already written in Schmidt normal form. The second
solution is given by solving $\tan\theta$ from dividing the first equation by
the latter on both sides. We then arrive at
\begin{equation}\label{ghz.sol}
\tan\theta=\frac{r_0}{2t},\quad \mu_2=\frac{hr_0+4t^2}{\sqrt{r^2_0+4t^2}},
\end{equation}
where $r_0=h+\sqrt{h^2+8t^2-4gt}$. This solution gives rise to a
basis $\{|q\rangle,\,|p\rangle\}$ defined as follows:
\begin{equation}\label{ghz.bas}
|q\ra=\frac{2t|0\ra+r_0|1\ra}{\sqrt{r^2_0+4t^2}},\quad
|p\ra=\frac{r_0|0\ra-2t|1\ra}{\sqrt{r^2_0+4t^2}}.
\end{equation}

To obtain the maximal entanglement we require that $\mu_1=\mu_2$. Because
$|qqq\ra$ is another, equally good dominant eigenvector, one can construct a
new Schmidt decomposition of $|\psi_0\ra$ whose factorizable basis consists of
$\{|q\rangle,\,|p\rangle\}$ instead of $\{|0\ra,|1\ra\}$ . This new
decomposition must not differ from the original one, since coefficients of the
canonical form are uniquely defined by state parameters. This in turn means
that the basis defined by $\{|q\rangle,\,|p\rangle\}$ results in a Schmidt
form equivalent to Eq.~(\ref{ghz.gsd}):
\begin{equation}\label{ghz.shm}
|\psi_0\ra=g|qqq\ra+t\left(|qpp\ra+|pqp\ra+
|ppq\ra\right)+h|ppp\ra.
\end{equation}
By expanding the l.h.s. of this equality in the computation basis
$\{0,1\}$ and identifying the corresponding coefficients with those in
Eq.~(\ref{ghz.gsd}), we arrive at the following three conditions on state
parameters
\begin{equation}\label{ghz.3con}
\frac{hr_0+4t^2}{\sqrt{r^2_0+4t^2}}=g,\;
\frac{2t(g-t)}{\sqrt{r^2_0+4t^2}}=t,\;
\frac{4t^2(r_0-h)}{\sqrt{r^2_0+4t^2}}=t(2h-r_0).
\end{equation}
By taking the ratios of the first two equations and the latter two equations,
and then eliminating $r_0$, we obtain the following single condition on state
parameters,
\begin{equation}\label{ghz.deg}
gh^2=(g+t)^2(g-2t).
\end{equation}
We remark that Eq.~(\ref{ghz.deg}) uniquely solves Eq.~(\ref{ghz.3con}) and is
in fact the degeneracy condition that forces the correct Schmidt
decomposition.

Let us rewrite the degeneracy condition in the following form
$g(g^2-h^2-3t^2)=2t^3$. Since $t\geq0$ it follows that $g^2\geq h^2+3t^2$.
Then from the normalization condition $g^2+h^2+3t^2=1$, it further gives that
$g^2\geq1/2$ and the lower bound $g^2=1/2$ is reached at $t=0, g=h$. The
resulting maximal entangled state is the celebrated GHZ state
\begin{equation}\label{ghz.ghz}
|GHZ\ra=\frac{|000\ra+|111\ra}{\sqrt{2}}.
\end{equation}

\section{W state}
Consider now the case $\gamma=\pi/2$. This case includes all W-class states.
Indeed, the three-tangle of a generic state Eq.~(\ref{int.gsd}) is given by
\begin{equation}\label{w.tau}
\tau=4g\sqrt{g^2h^4+16t_1^2t_2^2t_3^2+8gh^2t_1t_2t_3\cos2\gamma}.
\end{equation}
It vanishes if either
\begin{equation}\label{w.bisep}
gh^2=0 \ \mbox{and}\, t_1t_2t_3=0
\end{equation}
or
\begin{equation}\label{w.tau=0}
gh^2=4t_1t_2t_3\ne 0\ \mbox{and} \,\gamma=\pm\pi/2.
\end{equation}
The states satisfying Eq.~(\ref{w.bisep}) are bi-separable, namely, separable
with respect to A:BC, B:AC, or C:AB bipartition~\cite{vid-cir,gsd}, and the
maximally entangled states are Bell states between BC, AC or AB.

In Eq.~(\ref{w.tau=0}) the case of $\gamma=-\pi/2$ is equivalent to that of
$\gamma=\pi/2$ because the period of the angle $\gamma$ is $\pi$ and the point
$-\pi/2$ should be identified with $\pi/2$. All states satisfying
Eq.~(\ref{w.tau=0}) are W-class states~\cite{vid-cir} and conversely any
W-class state has a Schmidt decomposition with coefficients
Eq.~(\ref{w.tau=0}). Thus generic pure three-qubit states have 5 independent
real parameters, whiles W-class states have 3 of them.

Stationarity equations give four relevant solutions and one of them is
symmetric (under qubit permutations) while the remaining three others are not.
The easiest way for finding the symmetric solution $|qqq\ra$ is to set
$|q\ra=e^{i\pi/3}(\cos\theta|0\ra+i\sin\theta|1\ra)$ and solve the
stationarity equation~(\ref{int.stat}), just like what we did for the GHZ
case. The solution is
\begin{equation}\label{w.sol1}
\tan\theta=\frac{r_\pi}{2t},\quad
\mu=\frac{hr_\pi+4t^2}{\sqrt{r^2_\pi+4t^2}},
\end{equation}
where $r_\pi=h+\sqrt{h^2+8t^2+4gt}$. Since $r_\pi$ and $r_0$ differ only by
the sign of $t$, the degeneracy condition forcing $g=\mu$ can be obtained by
taking $t$ to $-t$ in Eq.~(\ref{ghz.deg}):
\begin{equation}\label{w.deg}
gh^2=(g-t)^2(g+2t).
\end{equation}

In the case of GHZ, the degeneracy condition is sufficient for finding the
maximally entangled state, but in the present case we need one more condition
by examining three other relevant solutions of stationarity equations. The
first solution is symmetric under the permutation of qubits A and B, but
asymmetric under other permutations and has a form $|qqq^\prime\ra$. Other two
solutions, with symmetric qubit pairs (A,C) and (B,C) respectively, give the
same eigenvalue. Thus, it suffices to consider the first solution of these in
addition to that in Eq.~(\ref{w.sol1}). Its constituent state $|q\ra$, up to
an irrelevant phase factor, can be parameterized by two angles, that is
$|q\ra=\cos\theta|0\ra+e^{i\vp}\sin\theta|1\ra$. (Similar parametrization can
be used for $|q'\ra$.) Straightforward but tedious algebra from solving the
stationarity conditions~(\ref{int.stat}) gives rise to the equations obeyed by
angles $\theta$ and $\vp$
\begin{equation}\label{w.asbloch}
\cos2\theta=\frac{h^2+gt-g^2}{h^2+g^2-3gt},\quad
\sin\varphi=\frac{h}{2g}\tan\theta,
\end{equation}
as well as the corresponding eigenvalue $\mu^\prime$ being
\begin{equation}\label{w.aseig}
{\mu^\prime}^2=\frac{g^2h^2-4gt^3}{g^2+h^2-3gt}.
\end{equation}
We shall present in the Appendix a simpler derivation using another approach
outlined in the next section.

Thus far we have obtained two different eigenvalues, namely $\mu$ and
$\mu^\prime$, and consequently $g$ has two different lower bounds. As argued
previously, for the maximally entangled state both lower bounds should be
saturated and this in turn uniquely defines state parameters. Indeed, by using
$\mu^\prime=g$ it follows that $(g+t)(g-2t)^2=0$ and hence $g=2t$ (as $g,t\ge
0$). Then the degeneracy condition~(\ref{w.deg}) forces $h^2=2t^2$. These two
conditions together with the normalization condition give $g=2/3$, $t=1/3$,
and $h=\sqrt{2}/3$ and thus yield the following maximally entangled state for
$\gamma=\pi/2$
\begin{equation}\label{w.max}
|\psi_{\pi/2}\ra=\frac{2}{3}|000\ra+\frac{1}{3}(|011\ra + |101\ra
+ |110\ra)+i\frac{\sqrt{2}}{3}|111\ra.
\end{equation}
This form turns out to be the generalized Schmidt normal form for the W-state.
Indeed, one can easily verify that $U\otimes U\otimes
U|\psi_{\pi/2}\ra=|W\ra$, where
\begin{equation}\label{w.w}
|W\ra=\frac{|100\ra+|010\ra+|001\ra}{\sqrt{3}},\quad
U=\frac{1}{\sqrt{3}}
\begin{pmatrix}
\sqrt{2} & -i\\
1 & i\sqrt{2}
\end{pmatrix}.
\end{equation}
Thus, W state is the maximally entangled state for $\gamma=\pi/2$. It should
be remarked that at $g=2t$ and $h^2=2t^2$ the fraction defining the angle
$\theta$ in Eq.~(\ref{w.asbloch}) is indefinite since both expressions in
denominator and numerator vanish. The reason is that the state
$|\psi_{\pi/2}\ra$(as well as the W-state) is an exceptional
state~\cite{preeti-09} and has countless nearest product states defined solely
by the condition $\tan\theta=2\sqrt{2}\sin\varphi$~\cite{tetrahedron}. All of
these product states are equally distant from $|\psi_{\pi/2}\ra$ (i.e.,
infinitely degenerate) and form a circle around it. This infinite degeneracy
is best captured when viewed in the computational basis. To be more precise,
the closest product states to W-state were previously shown~\cite{erw} to be
of the form
\begin{equation}
\left(e^{i\chi/2}\sqrt{\frac{2}{3}}|0\ra+e^{-i\chi}\sqrt{\frac{1}{3}}|1\ra\right)^{\otimes
3},
\end{equation}
where the arbitrariness of the phase $\chi$ clearly shows the infinite
degeneracy.

The eigenvalue $\mu^\prime$ also exhibits interesting features. If $g=2t$,
then $\mu^\prime=g$ and thus the solution Eq.~(\ref{w.asbloch}) maximizes the
overlap. On the other hand, if $gh^2=4t^3$ but $g\neq2t$, then the solution
minimizes the overlap, i.e., resulting in $\mu^\prime=0$. In order to obtain
further insight into this, we relates $\mu^\prime$ to the three-tangle $\tau$.
For the states in question $\tau=4g(gh^2-4t^3)$ and $\mu^\prime$ can be
written as
\begin{equation}\label{w.mu-tau}
{\mu^\prime}^2=\frac{g^2\tau}{\tau+4g(g+t)(g-2t)^2}.
\end{equation}
This shows clearly when  the asymmetric eigenvalue $\mu'$ takes the maximal
value and the minimal value.

\section{General case}\label{sec.gen}
Having warmed up by the previous two examples, we consider now the general case.
The GHZ and W are two special cases of the following treatment.

\subsection{Derivation of the degeneracy condition}
For symmetric states any solution of stationarity equations is symmetric under
the permutation of either the qubit pair AB, or AC or BC~\cite{tri}. Without
loss of generality we consider a solution that contains a symmetric pair A and
B, i.e. the product state has a form $|qqq^\prime\ra$ which, of course, does
not exclude the possibility $|q^\prime\ra=|q\ra$. Furthermore, theorem\,1 of
Ref.~\cite{reduced} states that the maximal overlap is uniquely determined
even if one party of the global pure state is traced out. This enables us to
express the maximal overlap in terms of the reduced density matrix $\rho^{AB}$
as follows
\begin{equation}\label{gen.red}
g^2=\max_{\varrho^1,\varrho^2}\tr\left(\rho^{AB}\varrho^1\otimes\varrho^2\right),
\end{equation}
where $\varrho^1$ and $\varrho^2$ are single-system pure state densities. The
density matrix $\rho^{AB}$ can be expanded in terms of identity operator and
the Pauli matrices $\sigma$'s,
\begin{equation}
\rho^{AB}=\frac{1}{4}\Big(\openone\otimes\openone+\rv\cdot{\bm\sigma}\otimes\openone
+\rv\cdot\openone\otimes{\bm\sigma}+{\bm\sigma}\cdot G\cdot{\bm\sigma} \Big),
\end{equation}
where $\rv$ is the Bloch vector of the qubit A(B) and the correlation matrix
$G$ is defined by the formula $G_{ij}=\tr(\rho^{AB}\sigma_i\otimes\sigma_j)$.
Explicitly
\begin{equation}\label{gen.bl}
\rv=(2ht\cos\gamma, 2ht\sin\gamma, g^2-h^2-t^2)
\end{equation}
and
\begin{equation}\label{gen.cor}
G=
\begin{pmatrix}
2t^2+2gt & 0 & -2ht\cos\gamma\\
0 & 2t^2-2gt & -2ht\sin\gamma\\
-2ht\cos\gamma & -2ht\sin\gamma & g^2+h^2-t^2
\end{pmatrix}.
\end{equation}

As $\rho^{AB}$ is symmetric under permuting parties, one can set
$\varrho^1=\varrho^2=\varrho$~\cite{wei-guh}. Denote by $\uv$ the
Bloch vector of the density matrix $\varrho$ that gives rise to
the maximum of $g^2$, then Eq.~(\ref{gen.red}) can be rewritten as
\begin{equation}\label{gam.g}
g^2=\frac{1}{4}(1+2\uv\cdot\rv+\uv\cdot G\uv).
\end{equation}
By introducing a Lagrange multiplier $\lambda$ that constraints
$\uv\cdot\uv=1$, the vector $\uv$ satisfies the following equation
\begin{equation}\label{gam.stat}
\rv+G\uv=\lambda\uv.
\end{equation}
Since $\lambda$ uniquely defines $\uv$ and $g$, two solutions have the same
eigenvalue if and only if they have the same Lagrange multiplier. By direct
substitution, it is easy to see that for the solution $u_z=1$ we have
$\lambda_0=2(g^2-t^2)$ and therefore the largest eigenvalue of
Eq.~(\ref{gam.stat}) is degenerate if there are two solutions at
$\lambda=\lambda_0$. Inserting this value into Eq.~(\ref{gam.stat}), one sees
that the necessary and sufficient condition for the existence of the second
solution corresponding to $\lambda=\lambda_0$ is ${\rm
det}(G-\lambda_0\openone)=0$, which can be rewritten as
\begin{equation}\label{gam.deg}
(g^2-t^2)^2(g^2-4t^2)=gh^2(g^3-3gt^2+2t^3\cos2\gamma).
\end{equation}
This is the degeneracy condition for arbitrary symmetric states
and the solutions contain all three-qubit states that can be
regarded as maximally entangled. When $\gamma=0$,
Eq.~(\ref{gam.deg}) reduces to Eq.~(\ref{ghz.deg}) and when
$\gamma=\pi/2$, it reduces either to Eq.~(\ref{w.deg}) or $g=2t$.
The latter is equivalent to Eq.~(\ref{w.aseig}).

\subsection{Maximally entangled three-qubit states.}
The degeneracy condition Eq.~(\ref{gam.deg}) can be considered as an algebraic
equation of degree six for $g$, where $t$ and $\gamma$ are free parameters and
$h^2$ can be further eliminated by the normalization condition
$h^2=1-g^2-3t^2$. It has six roots $g(t,\gamma)$ and as far as we are looking
for the largest eigenvalue, we should always take the largest root. In what
follows we will use the notations $g_1$ for the largest root and $g_2$ for the
second largest root. Equation~(\ref{gam.deg}) gives different types of lower
bounds depending on whether or not $\cos2\gamma$ is positive and below we
consider these two cases.

\begin{figure}
\includegraphics[width=7cm]{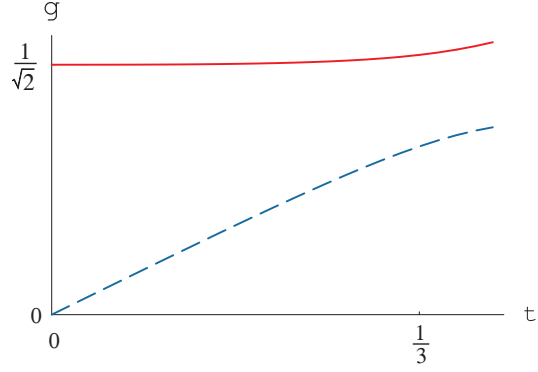}
\caption[fig1]{\label{fig:1}(Color online) Plots of
$t$-dependencies of the largest $g_1(t)$ (solid line) and  next
$g_2(t)$ (dashed line) roots of the degeneracy condition for
$\gamma=\pi/6$. $g_1(t)$ is the largest Schmidt coefficient and
has a minimum at $t=0$ which gives the GHZ-state.}
\end{figure}

Consider first states for which $0\leq\gamma\leq\pi/4$. The
degeneracy condition can be rewritten as follows
\begin{equation}\label{gam.ghz}
g^2(g^2-3t^2)(g^2-3t^2-h^2)=4t^6+2gh^2t^3\cos2\gamma.
\end{equation}
The right-hand side of this equation is positive and either $g^2\leq3t^2$ or
$g^2\geq3t^2+h^2$ holds. To find the maximal overlap we should take the latter
case
 $g^2\geq3t^2+h^2$ and then from the normalization condition it follows that
$g$ is minimal when $t=0$ and $g=h$. Thus the only maximally entangled state
in this class is the GHZ state: $(|000\rangle +
e^{i\gamma}|111\rangle)/\sqrt{2}$. It can be understood by reference to
Fig.~\ref{fig:1}, where $\gamma$ is taken to be $\pi/6$ for illustration. The
solid line represents the largest root $g_1(t)$ and the dashed line represents
the next root $g_2(t)$ as functions on $t$. It should be stressed that the
largest root $g_1(t)$ never intersects with other roots and therefore it is
the largest Schmidt coefficient for all values of $t$. Moreover, $g_1(t)$ is a
monotonically increasing function and our goal is to find its infimum (and
hence the supremum of the geometric measure), which is reached at $t=0$,
giving rise to the GHZ-state~\cite{open}.

\begin{figure}
\includegraphics[width=8cm]{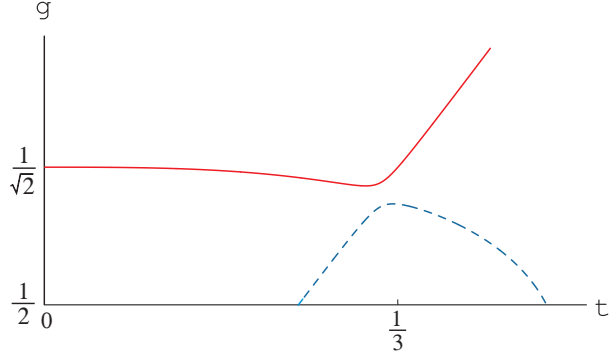}
\caption[fig2]{\label{fig:2}(Color online) Plots of
$t$-dependencies of the largest $g_1(t)$ (solid line) and  next
$g_2(t)$ (dashed line) roots of the degeneracy condition for
$\gamma=2\pi/5$. $g_1(t)$ has a minimum whiles $g_2(t)$ has a
maximum at $t=0.31943$. These roots never intersect unless
$\gamma=\pm\pi/2$.}
\end{figure}

Consider now the interval $\pi/4<\gamma<\pi/2$. Now the situation is different
since $g_1$ and $g_2$ are not monotonic functions. In Fig.~\ref{fig:2}
$g_1(t)$ (solid line) and $g_2(t)$ (dashed line) are plotted at
$\gamma=2\pi/5$. The largest root $g_1(t)$ has a minimum whiles the next root
$g_2(t)$ has a maximum at $t=0.31943$. These two roots never intersect unless
$\gamma=\pi/2$. Therefore the minimum of the function $g_1(t)$ is the lower
bound of the largest Schmidt coefficient. When $\gamma$ increases, the minimum
of $g_1$ moves towards larger $\gamma$ and decreases. Concurrently, the
maximum of $g_2$ moves towards larger $\gamma$ and increases. Thus, in the
range $\pi/4<\gamma<\pi/2$, one needs to search for the minimum value of
$g_1(t)$ over the allowable range of $t$, in order to find the maximal
entangled states.

\begin{figure}
\includegraphics[width=8cm]{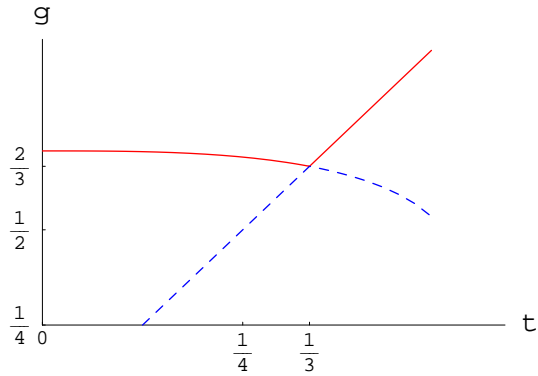}
\caption[figW]{\label{fig:W}(Color online) Plots of $t$-dependencies of the
largest $g_1(t)$ (solid line) and  next $g_2(t)$ (dashed line) roots of the
degeneracy condition for $\gamma=\pi/2$. The two curves touch at $(1/3,2/3)$.}
\end{figure}

 When $\gamma=\pi/2$ minimum of $g_1$ and maximum of $g_2$ coincide
at $t=1/3$ and this minimum value of $g_1$ yields the W-state. This is
illustrated in Fig.~\ref{fig:W}.

\begin{figure}
\includegraphics[width=8cm]{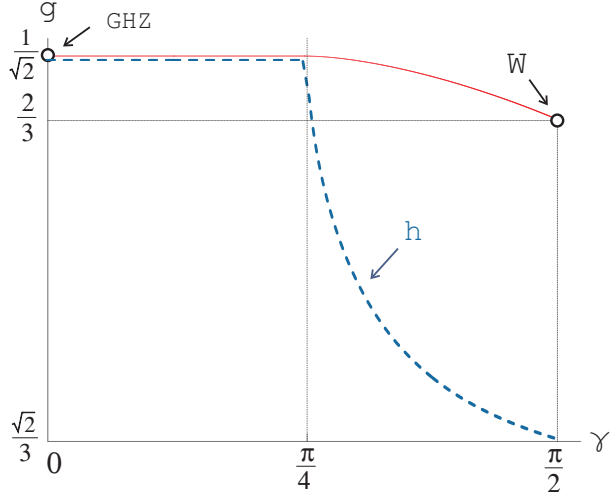}
\caption[fig3]{\label{fig:3}(Color online) The gauge phase $\gamma$-dependence
of the largest Schmidt coefficient $g$ as well as $h$ of maximally entangled
states. $g$ is constant and equal to $1/\sqrt{2}$ within
$0\leq\gamma\leq\pi/4$, then it decreases monotonically and becomes $2/3$ at
$\gamma=\pi/2$. Similarly, $h$ is constant and equal to $1/\sqrt{2}$ within
$0\leq\gamma\leq\pi/4$, then it decreases monotonically and becomes
$\sqrt{2}/3$ at $\gamma=\pi/2$. The parameter $t$ is obtained from the
normalization $g^2+3t^2+h^2=1$.}
\end{figure}

\begin{figure}
\includegraphics[width=8cm]{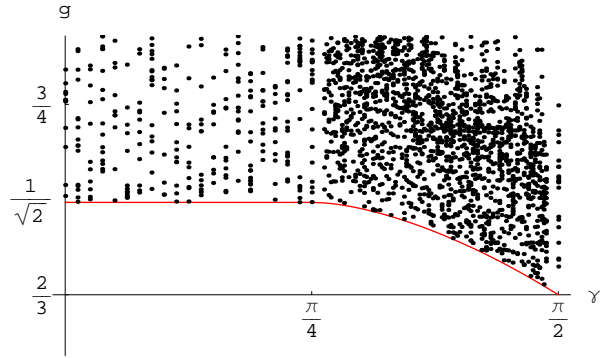}
\caption[fig_random]{\label{fig:random}(Color online) The maximal
overlap $g$ vs.~the gauge phase $\gamma$ for the family of
maximally entangled three-qubit states (red solid curve) as well
as randomly generated states (dots). This shows that the family of
states we have derived are indeed maximally entangled (with
minimal overlap $g$).  }
\end{figure}
We remark that the inequality $\min_t g_1(t)>\max_t g_2(t)$ holds for fixed
$\gamma\ne \pi/2$. This feature is verified numerically for all values
$0\leq\gamma<\pi/2$. This means that, given a fixed $\gamma$, the minimum of
the function $g_1(t)$ is the lower bound on $g$ and also provides
justification for the parametrization of maximally entangled states by the
gauge phase. The minimum of $g_1$ gives the value of the maximal overlap $g$
for the maximally entangled state at a given $\gamma$. Together with the value
of $t$ where the minimum of $g_1$ is achieved, the complete description of the
maximally entangled state is obtained (as $h$ is determined via the
normalization condition).

In Figure~\ref{fig:3} we shows the dependence of the $g$ and $h$ on the gauge
phase $\gamma$. The dependence of $t$ on $\gamma$ can be inferred from the
normalization $g^2+3t^2+h^2=1$. This then defines the family of the maximally
entangled three-qubit states. The parameters $g$ and $h$ are both constant and
equal to $1/\sqrt{2}$ within the interval $0\leq\gamma\leq\pi/4$, then
decrease with $\gamma$, with $g$ reaching 2/3 and $h$ reaching $\sqrt{2}/3$ at
$\gamma=\pi/2$. We remark that we have assumed the maximally entangled states
have the permutation invariant form~(\ref{int.gsd2}), and this is supported by
our numerical test that states generated randomly do not achieve  $g$ below
(or entanglement above) those of the maximally entangled states; see
Fig.~\ref{fig:random}.

\subsection{Survey of maximally entangled states}

\begin{figure}
\includegraphics[width=8cm]{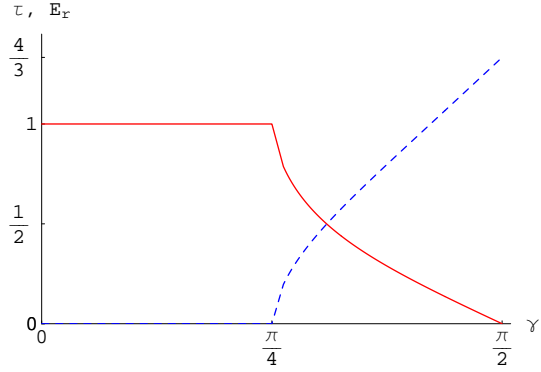}
\caption[fig5]{\label{fig:5}(Color online) The phase dependence of
three-tangle $\tau$ (red solid curve) and the residual bipartite entanglement
$E_r$ (blue dashed curve) vs.~$\gamma$ for the family of maximally entangled
three-qubit states. }
\end{figure}

\begin{figure}
\includegraphics[width=8cm]{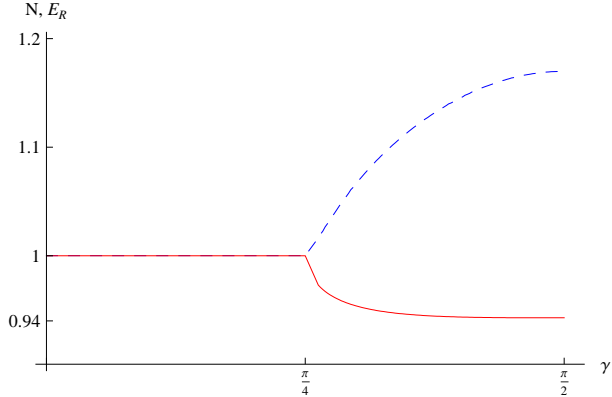}
\caption[figner]{\label{fig:ner}(Color online) The phase dependence of the
negativity $N$ (red solid curve) and the relative entropy of entanglement
$E_R$ (blue dashed curve) vs.~$\gamma$ for the family of maximally entangled
three-qubit states. }
\end{figure}

Once the maximally entangled three-qubit states (and hence their maximal
overlap) have been obtained as a function of the gauge phase $\gamma$, it is
of interest to compare the results with other measures. Two other quantities
of relevance are the aforementioned three-tangle $\tau$~\cite{coff} and the
residual bipartite entanglement $E_r$~\cite{w}. We show the
$\gamma$-dependence of $\tau$ and $E_r$ for the maximally entangled
three-qubit states in Fig.~\ref{fig:5}. We also compare the
$\gamma$-dependence of yet two other measures, the bi-partition negativity
($N$) and the relative entropy of entanglement ($E_R$) in Fig.~\ref{fig:ner}.
The three-tangle $\tau$ and the negativity $N$ decrease monotonically for
$\gamma\in[\pi/4,\pi/2]$, whereas the residual bipartite entanglement $E_r$
and the relative entropy of entanglement $E_R$ increase and achieve $E_r=4/3$
and $E_R=\log_2 (9/4)$ at $\gamma=\pi/2$ for the W-state. The geometric
measure for the maximally entangled states is equal to $1-g^2$, and hence, as
can be inferred from Fig.~\ref{fig:3}, it increases monotonically. To
summarize the behaviors of three different measures, we have that, in going
from GHZ to W as $\gamma$ increases, the geometric measure and the relative
entropy of entanglement and the bipartite entanglement all increase
monotonically whereas the three-tangle and the negativity both decrease
monotonically.

Instead of the geometric measure, one may well use other entanglement measures
to derive the maximally entangled states. Conversely, having at one's disposal
the set of maximally entangled states one can analyze and compare different
entanglement measures. In this view the behaviors of the different measures
can be understood as follows. The three-tangle quantifies genuine tripartite,
i.e. GHZ-type, entanglement, but does not detect W-type entanglement at
all~\cite{footnote}. Then all states within the interval
$-\pi/4\leq\gamma\leq\pi/4$ possess only GHZ-type entanglement and in going
away from these states, the three-tangle detects the residual of GHZ-type
entanglement. In this regards it decreases with $\gamma$ from $\tau=1$ at
GHZ-state and vanishes at W-state. On the contrary, the residual bipartite
entanglement quantifies the W-type entanglement and does not detect the
GHZ-type entanglement at all. Hence its behavior is opposite to that of the
three-tangle. The negativity quantifies entanglement across bi-partition. As
the GHZ state is equivalent to a Bell state if one makes the bi-partition
A:BC, it possesses the largest negativity $N=1$. The W state possesses less
negativity $N=2\sqrt{2}/3$. And the family becomes less and less similar to
GHZ as one increases the gauge phase, one expects a gradual interpolation of
the negativity between these values. The situation of the geometric measure is
very different from these measures. It quantifies whole entanglement present
in the state and, owing to this, detects an one-parameter set of maximally
entangled states. The behavior of the relative entropy of entanglement for
this family is qualitatively similar to that of the geometric measure, and
this is expected as the geometric measure can be used to provide lower bounds
on the relative entropy of entanglement~\cite{erw}, with GHZ and W states
saturating the bounds.

It is interesting to note that the manifold of GHZ-class maximally entangled
states is an open manifold in a sense that there is no state within GHZ-class
states that has the global maximal geometric measure of entanglement (which is
actually possessed by the W-state). Indeed, let us analyze $\gamma=\pi/2$ case
once again. The three-tangle $\tau$ of these states is given by the formula
$\tau=4g(gh^2-4t^3)$ and the degeneracy condition Eq.~(\ref{w.deg}) can be
rewritten as follows
\begin{equation}\label{sp.tau}
4g(g+t)^2(g-2t)=\tau.
\end{equation}
All of these states are GHZ-class state unless the limit $\tau=0$ (and hence
$g=2t$, which means that the asymmetric solution comes into action) is
reached. Near this limit $g$ depends on $\tau$ by the asymptotic formula
$g=2/3+\sqrt{3\tau/8}+O(\tau)$ which can be derived from Eq.(\ref{sp.tau}). On
the other hand, if $0\leq\gamma<\pi/2$ then $g>2/3$. Thus the largest Schmidt
coefficient of the GHZ-class states comes arbitrarily close to $2/3$ but never
reaches it as the lower bound is only achieved by the W state.
What about the interconversion between states in the family of the maximally
entangled states? We have seen that in going from GHZ to W the geometric
measure of entanglement and the bipartite entanglement both increase
monotonically whereas the three-tangle decreases monotonically (see
Fig.~\ref{fig:5}). The two different trends of these entanglement monotones
for this family of states imply that deterministic interconversion via LOCC
among these maximally entangled three-qubit states is not possible. The
results of the three-tangle $\tau$ and the D\"ur-Vidal-Cirac classification of
entangled states~\cite{vid-cir} show that under stochastic LOCC, any pure
states can be probabilistically converted to one another within the same
class, and that there are five classes of three-qubit pure states: one class
of completely separable states, three classes of biseparable states, GHZ-class
and W-class states. But states in one class cannot be converted to any other
in a different class even with SLOCC. In our case, the whole family of the
maximally entangled three-qubit states, except the W state (which has
$\tau=0$), belong to the GHZ-class (with $\tau>0$). Therefore, under SLOCC,
any states in the family of maximally entangled three-qubit states can be
probabilistically converted to one another in the family except to/from the W
state.
\section{Concluding remarks}
 We use a generalized Schmidt decomposition and the
geometric measure of entanglement to characterize three-qubit pure states and
derive a single-parameter (characterized by the gauge phase) family of
maximally entangled three-qubit states. The resulting family of maximally
entangled states connect continuously from GHZ to W state. The paradigmatic
Greenberger-Horne-Zeilinger (GHZ) and W states emerge as extreme members in
this family of maximally entangled states.

This family of maximally entangled states turn out to possess interesting
features of entanglement. In going from GHZ to W the geometric measure and the
relative entropy of entanglement and the bipartite entanglement all increase
monotonically whereas the three-tangle and the negativity both decrease
monotonically. { This clearly exemplifies the ordering issue in the
multipartite entanglement.}
 It also
implies that deterministic interconversion via LOCC among these maximally
entangled three-qubit states is not possible. However, the results of the
three-tangle $\tau$ and the D\"ur-Vidal-Cirac classification of entangled
states show that under {\it stochastic\/} LOCC, any states in the family of
maximally entangled three-qubit states can be probabilistically converted to
one another in the family (except to/from the W state).

 In general, three-qubit pure states require 14 real independent
parameters to completely characterize. The use of local-unitary equivalence
helps to reduce the number of parameters necessary for the characterization of
entangled states. Via the generalized Schmidt decomposition, we have navigated
through the remaining vast space and identified the one-parameter maximally
entangled states via the geometric measure. But one may as well use other
entanglement measures. Will other measures give rise to such a non-trivial
family of states? For example, using the gauge phase as the parametrization
and the three-tangle as the characterization of entanglement, one obtains that
the maximally entangled states being $(|000\rangle +
e^{i\gamma}|111\rangle)/\sqrt{2}$, which are essentially the same GHZ state.
Such featureless family of states also arise when the characterization of
entanglement is replaced by negativity. On the other hand, when using the
relative entropy of entanglement, one expects that the resulting family of
maximally entangled states will have similar feature to those via the
geometric measure, albeit not identical. (It is because that the geometric
measure serves as a lower bound of the relative entropy of entanglement.)
Perhaps the procedure that we have gone through to identify the maximally
entangled states can serve as two purposes: (1) to explore certain cross
section of Hilbert space
 via suitable parametrization and choice of
entanglement measures; (2) to investigate and compare the behavior of
entanglement measures via the resulting family of maximally entangled states:
whether they are interesting or featureless. In doing so, one may identify
certain distinct states or classify various measures of entanglement. In the
former respect, it remains to be seen whether the derived family of entangled
states can be of any use in quantum information processing tasks previously
unexplored. In the latter respect, one thus has that the various entanglement
measures discussed in the present paper can be divided into two different
groups: (a) the geometric measure of entanglement, the relative entropy of
entanglement, and the residual bi-partite entanglement; (b) the three-tangle
and the negativity.

 While the nonlinear eigenvalues of the W state are infinitely
degenerate, eigenvalues of the GHZ-class states are doubly degenerate and thus
no invertible local operations can match them. As a remark, it is interesting
to see that the whole W-class states have only a single representative in the
above one-parameter family of maximally entangled states, namely, the W state,
which can be regarded geometrically as the center of the largest full-sphere
with no unentangled states.

The W-class states form a boundary set for pure three-qubit states and the
boundary is the limit $\gamma\to\pi/2,\;gh^2\to4t_1t_2t_3$. Accordingly, the
W-state is the the endpoint of maximally entangled pure states~\cite{pure-05}.
The boundary behavior of entanglement is different and owing to this the set
of GHZ states is noncompact. Hence GHZ-class states should have noncountably
infinite collection of maximally entangled states approaching to the W-state.

 Several outstanding questions remain: does the minimal reversible entanglement
 generating set exist? If so, what does it consist of? Can the family of
 states derived in the present paper constitute in part the generating set?
 As the consideration of such questions involve reversible conversion of states in
 the asymptotic limit, few progress has been made. Unfortunately, in the
 present paper such questions are not answered and remain open.

\begin{acknowledgments}
ST thanks Anthony Sudbery  and Levon Tamaryan for stimulating
discussions. TCW is supported by IQC, NSERC and ORF. DKP is
supported by the National Research Foundation
(F01-2008-000-10039-0).
\end{acknowledgments}

\appendix
\section{Solutions of stationarity equations}

As indicated above, we have found all six solutions of stationarity equations
for $\gamma=0$ and $\gamma=\pi/2$. In general these equations have two types
of solutions: the nearest separable states for highly and slightly entangled
three qubit states, respectively. There are no slightly entangled states
within the subclass of symmetric states and corresponding solutions are
irrelevant since $\mu<g$ and the equality $\mu=g$ is never saturated.

At $\gamma=0$ the stationarity equation~(\ref{gam.stat}) along
the axis $y$ reduces to
\begin{equation}\label{ap.fac0}
(2gt-2t^2)u_y=\lambda u_y
\end{equation}
Either $u_y=0$ or $\lambda=2t^2-2gt$. The second case does not give a true
maximum since $\lambda<0$.

Similar situation occurs at $\gamma=\pi/2$. The stationarity
equation~(\ref{gam.stat}) along the axis $x$ reduces to
\begin{equation}\label{ap.facpi}
(2t^2+2gt)u_x=\lambda u_x
\end{equation}
The case $u_x=0$ gives three symmetric solutions, while
the remaining case $\lambda=2t^2+2gt$ gives three asymmetric
solutions.

Thus at extreme values of the gauge phase stationarity equations are factorized
to cubic equations.  One of the roots of these cubic equations is either
$\sin\theta_k=0$ or $\tan\theta_k=\pm(g\pm t)/h$, where the angle $\theta_k$
defines the weights of computational basis vectors in the local state as
follows $|i_k\ra\sim\cos\theta_k|0\ra+e^{i\vp}\sin\theta_k|1\ra$. Hence we know
one root of each cubic equation and we can find the other two by solving a
quadratic equation. In this way we find all roots of the characteristic polynomial.

\paragraph{Solutions in the case of $\gamma=0$.}
We list them as follows.\\
 Solution 1 (symmetric, standard)
$$|q_1q_2q_3\ra=|000\ra,\quad \mu=g.$$
Solution 2 (symmetric, relevant)
$$|q_1q_2q_3\ra=|qqq\ra,\quad
|q\ra=\frac{2t|0\ra+r_+|1\ra}{\sqrt{r_+^2+4t^2}},\quad
\mu=\frac{hr_++4t^2}{\sqrt{r_+^2+4t^2}},\quad
r_+=h+\sqrt{h^2+8t^2-4gt}.$$
 Solution 3 (symmetric, irrelevant)
$$|q_1q_2q_3\ra=|qqq\ra,\quad
|q\ra=\frac{2t|0\ra+r_-|1\ra}{\sqrt{r_-^2+4t^2}},\quad
\mu=\frac{hr_-+4t^2}{\sqrt{r_-^2+4t^2}},\quad
r_-=h-\sqrt{h^2+8t^2-4gt}.$$
 Solution 4 (asymmetric, irrelevant)
$$|q_1q_2q_3\ra=|qqq^\prime\ra,\quad
|q\ra=\frac{h|0\ra-(g+t)|1\ra}{\sqrt{h^2+(g+t)^2}}, \quad
\mu^2=\frac{g^2h^2+t^2(g+t)^2}{h^2+(g+t)^2}, \quad
|q^\prime\ra=\frac{\la qq|\psi_0\ra}{\mu}.$$
 Solution 5 (asymmetric, irrelevant, permutation of fourth
solution)
$$|q_1q_2q_3\ra=|qq^\prime q\ra.$$
 Solution 6 (asymmetric, irrelevant, permutation of fourth
solution)
$$|q_1q_2q_3\ra=|q^\prime qq\ra.$$

\smallskip

\paragraph{Solutions in the case of $\gamma=\pi/2$.} Solutions for
$\gamma=\pi/2$ are:\\
Solution 1 (symmetric, standard)
$$|q_1q_2q_3\ra=|000\ra,\quad \mu=g,$$
Solution 2 (symmetric, relevant)
$$|q_1q_2q_3\ra=|qqq\ra,\quad |q\ra =
e^{i\pi/3}\frac{2t|0\ra+ir_\pi|1\ra}{\sqrt{r_\pi^2+4t^2}},
\quad\mu=\frac{hr_\pi+4t^2}{\sqrt{r_\pi^2+4t^2}},\quad
r_\pi=\sqrt{h^2+4gt+8t^2}+h.$$
Solution 3 (symmetric, irrelevant)
$$|q_1q_2q_3\ra=|qqq\ra,\quad |q\ra =
e^{i\pi/3}\frac{2t|0\ra-is_\pi|1\ra}{\sqrt{s_\pi^2+4t^2}},
\quad\mu=\frac{hs_\pi-4t^2}{\sqrt{s_\pi^2+4t^2}},\quad
s_\pi=\sqrt{h^2+4gt+8t^2}-h.$$
Solution 4 (asymmetric, relevant)
\begin{eqnarray*}
 & &|q_1q_2q_3\ra=|qqq^\prime\ra,\quad|
q\ra=\cos\theta|0\ra+e^{i\vp}\sin\theta|1\ra,\quad
{\mu^\prime}^2=\frac{g^2h^2-4gt^3}{g^2+h^2-3gt},\\
 & &\cos2\theta=\frac{h^2+gt-g^2}{h^2+g^2-3gt},\quad
\sin\varphi=\frac{h}{2g}\tan\theta.
\end{eqnarray*}
Solution 5 (asymmetric, relevant, permutation of fourth
solution)
$$|q_1q_2q_3\ra=|qq^\prime q\ra.$$
Solution 6 (asymmetric, relevant, permutation of fourth
solution)
$$|q_1q_2q_3\ra=|q^\prime qq\ra.$$


\begin{thebibliography}{99}
\bibitem{minimum}These solutions are minimazing the overlap. Stationarity
equations do not differ maximums and mimimus and give both extremes.
Surprisingly, there is a solution which, depending on state parameters,
maximizes the overlap for some states and minimizes for some other states.
This solution is presented in section IV.
\bibitem{tetrahedron}Notice, the maximal angle $\theta_{\max}$ is given
by $\tan\theta_{\max}=2\sqrt{2}$ and this is the angle between sides of a
regular tetrahedron.
\bibitem{open}Notice, when $t=0$ the phase $\gamma$ can be further
made zero by chaning local phases as $e^{i\gamma}$  can be
absorbed in the definition of the product state $|111\ra$. It
turns out that for $0\le\gamma\leq\pi/4$ the maximally
entanglement is achieved when $t=0$ and GHZ is the only maximally
entangled state in this range.
\bibitem{footnote}For mixed state the three tangle does not
properly quantify even some GHZ-type states~\cite{eylee09}.
\end{thebibliography}
\end{document}